\documentclass[trackchanges]{aastex701}

\begin{document}

\title{The Sound of Noise: Leveraging the Inductive Bias of Pre-trained Audio Transformers for Glitch Identification in LIGO}

\author[0009-0007-9657-1412]{Suyash Deshmukh}
\affiliation{Department of Physics and Astronomy, Vanderbilt University\\ 2201 West End Avenue, Nashville, Tennessee - 37235,}
\email[show]{suyash.deshmukh@vanderbilt.edu}

\author[0000-0001-8700-3455]{Chayan Chatterjee}
\affiliation{Department of Physics and Astronomy, Vanderbilt University\\ 2201 West End Avenue, Nashville, Tennessee - 37235,}
\affiliation{Data Science Institute, Vanderbilt University\\ 1400 18th Avenue South Building, Suite 2000, Nashville, Tennessee - 37212,}
\email[show]{chayan.chatterjee@vanderbilt.edu}

\author[0000-0002-7627-5444]{Abigail Petulante}
\affiliation{Data Science Institute, Vanderbilt University\\ 1400 18th Avenue South Building, Suite 2000, Nashville, Tennessee - 37212,}
\email[show]{abigail.petulante@vanderbilt.edu}

\author[0000-0002-1166-2005]{Tabata Aira Ferreira}
\affiliation{Instituto Nacional de Pesquisas Espaciais (INPE), São José dos Campos, SP, Brazil} 
\email[show]{tabata.aira@inpe.br}

\author[0000-0003-1007-8912]{Karan Jani}
\affiliation{Department of Physics and Astronomy, Vanderbilt University\\ 2201 West End Avenue, Nashville, Tennessee - 37235,} 
\email[show]{karan.jani@vanderbilt.edu}

\begin{abstract}
Transient noise artifacts, or glitches, fundamentally limit the sensitivity of gravitational-wave (GW) interferometers and can mimic true astrophysical signals, particularly the short-duration intermediate-mass black hole (IMBH) mergers. Current glitch classification methods, such as Gravity Spy, rely on supervised models trained from scratch using labeled datasets. These approaches suffer from a significant ``label bottleneck," requiring massive, expertly annotated datasets to achieve high accuracy and often struggling to generalize to new glitch morphologies or exotic GW signals encountered in observing runs. In this work, we present a novel cross-domain framework that treats GW strain data through the lens of audio processing. We utilize the Audio Spectrogram Transformer (AST), a model pre-trained on large-scale audio datasets, and adapt it to the GW domain. Instead of learning time-frequency features from scratch, our method exploits the strong inductive bias inherent in pre-trained audio models, transferring learned representations of natural sound to the characterization of detector noise and GW signals, including IMBHs. We validate this approach by analyzing strain data from the third (O3) and fourth (O4) observing runs of the LIGO detectors. We used t-Distributed Stochastic Neighbor Embedding (t-SNE), an unsupervised clustering technique, to visualize the AST-derived embeddings of signals and glitches, revealing well-separated groups that align closely with independently validated Gravity Spy glitch classes. Our results indicate that the inductive bias from audio pre-training allows superior feature extraction compared to traditional supervised techniques, offering a robust, data-efficient pathway for discovering new, anomalous transients, and classifying complex noise artifacts in the era of next-generation detectors.

\end{abstract}



\section{Introduction} 

The release of the fourth Gravitational-Wave Transient Catalog (GWTC-4) \citep{GWTC-4_1} marks a significant milestone in gravitational-wave (GW) astronomy. Covering the first half of the fourth observing run (O4a), the LIGO-Virgo-KAGRA (LVK) Collaboration \citep{LIGO, Virgo, KAGRA} reported 128 new confident GW candidates, bringing the cumulative total of confirmed detections to over 200 \citep{GWTC-4_1, GWTC-4_populations}. This dramatic increase in detection rate --averaging one event every few days -- is a testament to the enhanced sensitivity of the current detector network. However, one of the major challenges of GW data analysis is the highly nonstationary and non-Gaussian nature of the detector strain data, which is susceptible to contamination by noise transients, or ``glitches," that complicate the search for true astrophysical signals \citep{GW170817_glitch, glitches_definition1, glitches_definition2, glitches_definition3}. These short-duration noise artifacts, arising from environmental coupling and instrumental instabilities, present a formidable challenge to detector characterization and signal analysis \citep{Impact_of_glitch1, Impact_of_glitch2, Impact_of_glitch3, Impact_of_glitch4, Impact_of_glitch5, Impact_of_glitch6}. For example, during the third observing run (O3), approximately 25\% of all confident detections required manual noise mitigation due to overlapping noise artifacts \citep{GWTC-3, Glitch_overlapping_with_signal}. In particular, short-duration artifacts such as ``blip" glitches are known to mimic the morphology of high-mass binary black holes (BBHs), creating a risk of false alarms that can pollute population studies. Recent work has also shown that non-Gaussian noise artifacts can create biases in the analysis of GW events \citep{GW200129_4, GW200129_3, udall2025inferringspinsmergingblack}. \\

To combat this, the LVK Collaboration employs a suite of detector characterization tools. Low-latency pipelines like Omicron \citep{Omicron} utilize Q-transforms to identify excess power in the time-frequency domain. Statistical approaches such as iDQ \citep{iDQ} and HVeto \citep{HVeto} leverage auxiliary channels to veto data segments with high probabilities of environmental coupling. While effective for known instrumental noises, these tools often struggle with novel glitch classes that lack clear auxiliary witnesses. Consequently, machine learning has become indispensable in this space. Among the standard tools is Gravity Spy \citep{GravitySpy, GravitySpy_classifier, GravitySpy_attention}, a Convolutional Neural Networks (CNNs) trained on hundreds of thousands of citizen-science labels to categorize glitches into morphological classes. Building on this, GSpyNetTree \citep{GSpyNetTree} utilizes a decision tree of multilabel CNN classifiers, organized by the estimated mass of the candidate event, to robustly distinguish between astrophysical signals and morphologically similar glitches even when they overlap in time. \\

However, supervised classification is inherently reactive as it relies on extensive `known' and labeled datasets. As detectors change configurations between runs (e.g., from O3 to O4), glitch morphologies evolve, rendering models trained on previous data less effective and necessitating expensive retraining campaigns. To capture unknown glitch morphologies without the latency of manual labeling, unsupervised approaches are essential. Addressing this need, Ferreira and González \citep{Tabata_tSNE} used t-Distributed Stochastic Neighbor Embedding (t-SNE) on features derived directly from the Q-transform to visualize glitch populations in O3b data without the need for discrete labels, which can also be used to explore possible correlations with instrumental and environmental conditions~\citep{ferreira2025analysis}.\\

In this work, we propose a paradigm shift from training domain-specific models from scratch to adapting large-scale ``foundation models" pre-trained on vast audio datasets. The core hypothesis is that the spectro-temporal features learned from millions of hours of natural audio transfer effectively to glitch analysis. By leveraging a model that has already learned to process sound, we exploit a strong inductive bias that facilitates rapid learning and robust generalization with minimal labeled data. \\

\section{The Foundational AI Paradigm}

Foundation models represent a paradigm shift in Artificial Intelligence, characterized by large-scale neural networks trained on massive, diverse datasets using self-supervised or weakly supervised learning objectives. Unlike traditional models initialized for a single specific task, foundation models learn robust, generalized representations of data that can be adapted (or ``fine-tuned") to a wide array of downstream applications with minimal task-specific data. In the audio domain, modern transformer architectures are typically pre-trained on datasets comprising millions of structured audio clips (e.g., AudioSet \citep{AudioSet}). This is where the concept of inductive bias becomes crucial. Inductive bias refers to the set of assumptions a learning algorithm makes to predict outputs for inputs it has not encountered. It is essentially the prior knowledge built into the model that enables it to generalize from limited examples. Standard CNNs initialized from scratch possess a weak inductive bias regarding audio since they treat spectrograms as generic images and must learn to detect edges, curves, and chirps from scratch, requiring vast amounts of labeled data to discover features. In contrast, by leveraging a pre-trained audio transformer, we exploit a strong inductive bias: the model already knows how spectro-temporal features evolve, allowing it to characterize complex glitch morphologies as variations of known acoustic structures rather than abstract patterns of pixels. \\

Recent work has begun to explore this cross-domain potential. Chatterjee et al \citep{GW-Whisper}. introduced GW-Whisper, fine-tuning the Whisper \citep{Whisper} Automatic Speech Recognition (ASR) model for GW detection. Here, we employ the Audio Spectrogram Transformer (AST) \citep{AST}, a vision-transformer model that processes spectrograms as images. AST captures long-range dependencies and global semantic context more naturally than sequential models. We present a novel framework that adapts AST to LIGO O3 and O4 strain data by fine-tuning the model on glitch data and projecting the embeddings from the fine-tuned model into a lower-dimensional space. We extend the morphological analysis of Ferreira and González \citep{Tabata_tSNE} to demonstrate that the strong inductive bias of AST allows for superior clustering and discovery of novel glitch classes, offering a robust, data-efficient pathway for detector characterization. \\

\subsection{Audio Spectrogram Transformer}

AST is a pure attention-based model adapted from Vision Transformers (ViTs) for audio classification. Unlike CNNs that process spectrograms using localized sliding windows, AST treats the input spectrogram as a unified visual entity, enabling it to capture long-range global dependencies between distant time-frequency features. The model takes as input a log-mel spectrogram with dimensions $T \times F$, where $T$ represents time frames and $F$ represents frequency bins. To make this compatible with the Transformer architecture, the 2D spectrogram is split into a sequence of $N$ patches of size $16 \times 16$ pixels. Each patch is flattened into a 1D vector and linearly projected into an embedding space of dimension $D=768$. Since the standard self-attention mechanism is permutation-invariant (i.e., it does not inherently know the order of the patches), a learnable positional embedding is added to each patch embedding. This allows the model to retain crucial spatial information, distinguishing whether a frequency chirp is rising (up-chirp) or falling (down-chirp) over time. \\

The resulting sequence of patch embeddings is fed into the Transformer Encoder. The encoder consists of 12 stacked layers, each containing a Multi-Head Self-Attention (MSA) mechanism \citep{Transformer} and a Multi-Layer Perceptron (MLP) block. The self-attention mechanism allows every patch in the spectrogram to effectively attend to every other patch, regardless of their distance in time or frequency. In our workflow, we extract this 768-dimensional vector as the semantic embedding for downstream clustering and analysis. \\

\subsection{Fine-tuning Audio Spectrogram Transformer}

\begin{figure}
 \centering
        \includegraphics[width=0.7\textwidth]{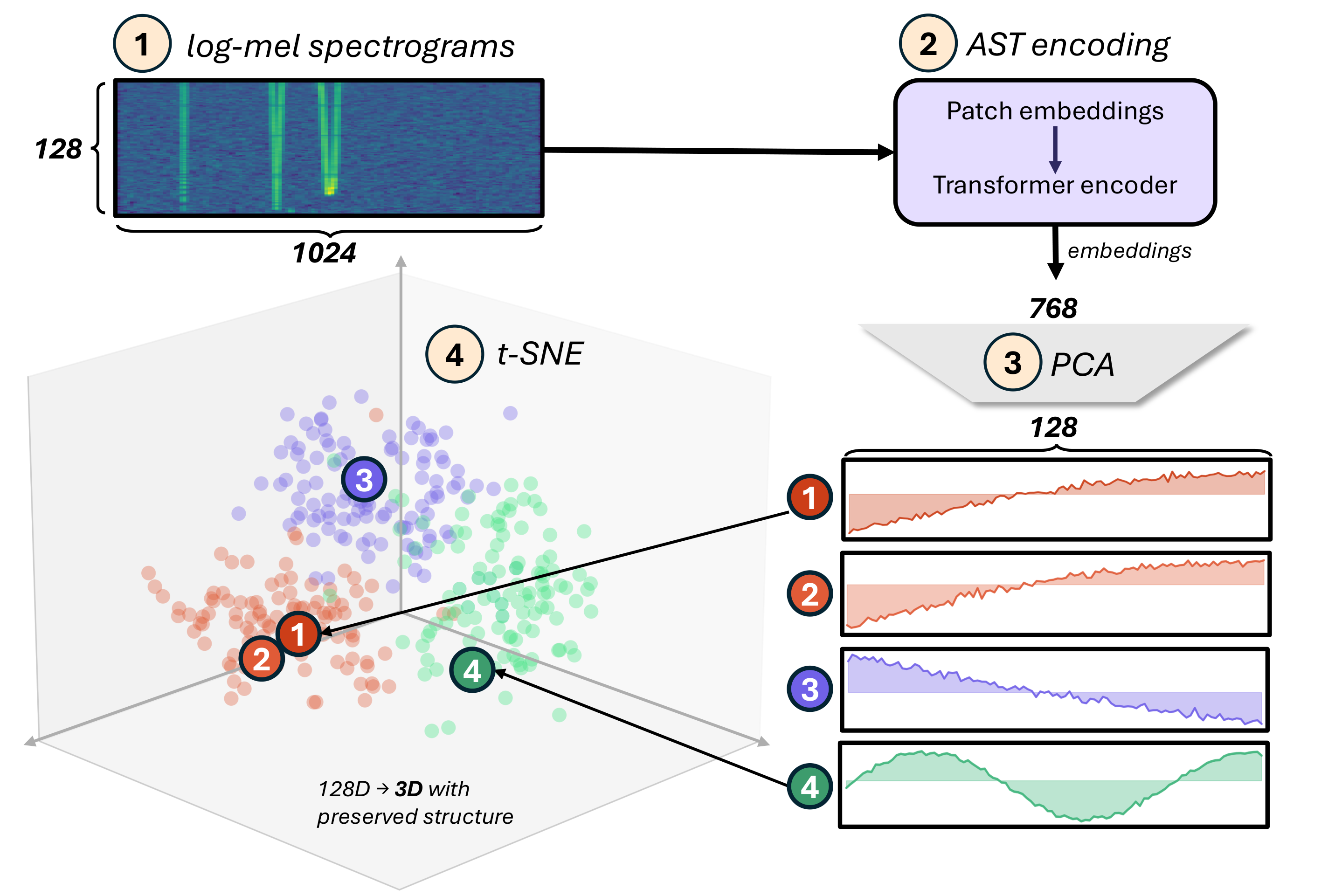}
 \caption{Overview of the proposed AST-LoRA glitch characterization workflow. (1) Input strain data is converted into log-mel spectrograms of shape ($1024 \times 128$). (2) The spectrograms are encoded by the AST encoder, fine-tuned via LoRA, to extract rich 768-dimensional semantic embeddings. (3) The high-dimensional embeddings undergo dimensionality reduction via PCA. (4) The reduced features are projected into a 3D latent space using t-SNE, revealing distinct clusters of glitch morphologies. The panels on the right display representative embedding vectors corresponding to the centroids of the identified clusters (colored red, orange, blue, and green).}
\label{fig:fig1}
\end{figure}

To adapt the representations learned from natural audio to the specific domain of GW detector noise, we employ Parameter-Efficient Fine-Tuning (PEFT) \citep{PEFT}. Standard fine-tuning, which updates all parameters of a large pre-trained model, is often computationally prohibitive and prone to catastrophic forgetting, where the model loses its robust prior knowledge in favor of overfitting to the smaller target dataset. PEFT addresses this by freezing the majority of the pre-trained parameters and updating only a small subset of additional weights, significantly reducing memory requirements while maintaining high performance. Specifically, we utilize Low-Rank Adaptation (LoRA) \citep{LoRA}. LoRA hypothesizes that the change in weights during model adaptation has a low ``intrinsic rank". Instead of updating a full-rank weight matrix $W$ directly, LoRA freezes $W$ and injects two trainable low-rank matrices, $A$ and $B$, such that the weight update is represented as $\Delta W = B \times A$. During the forward pass, the input is processed by both the frozen pre-trained weights and the low-rank adaptation branch, and their outputs are summed. This allows us to fine-tune AST on LIGO strain data with a fraction of the trainable parameters required for full fine-tuning, effectively steering the model's audio inductive bias toward glitch morphology without destroying its learned feature extractors. \\

\begin{figure}
 \centering
        \includegraphics[width=1.0\textwidth]{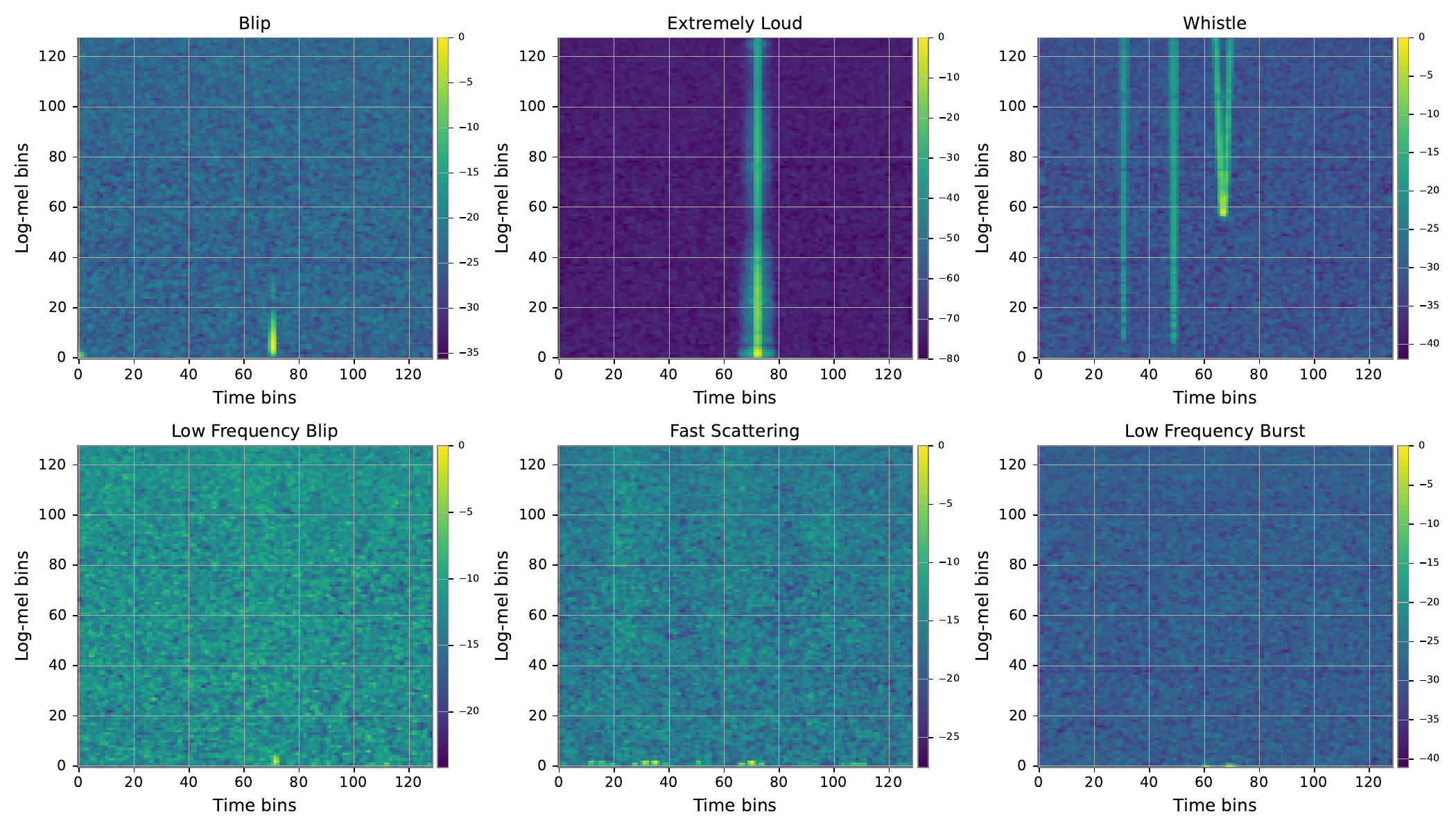}
 \caption{Log-mel spectrograms of a Blip, Extremely loud and Whistle glitch (top) and a Low frequency blip, Fast scattering and Low frequency burst glitch (bottom).}
\label{fig:fig2}
\end{figure}

Our complete analysis pipeline is illustrated in Fig.~\ref{fig:fig1}. The raw GW strain data is preprocessed into log-mel spectrograms in step 1. We generate spectrograms with 128 frequency bins and a temporal resolution resulting in a $1024 \times 128$ input representation, aligning with the input dimensions expected by the standard AST architecture.  Fig.~\ref{fig:fig2} shows the log-mel spectrograms of some common glitches in LIGO. The top panel shows Blip, Extremely loud and Whistle glitch which are easily distinguishable in log-mel spectrograms because of their time-frequency morphology and high SNR. The bottom panels shows Low frequency blip, Fast scattering and Low frequency burst which are harder to distinguish visually because of their confinement to lower frequency bins and lack of distinct spectro-temporal features. These spectrograms are passed to the AST encoder (Step 2), which splits the 2D input into a sequence of $16 \times 16$ patches. These patches are linearly projected, combined with positional embeddings, and processed through the Transformer layers. It is within these layers that the LoRA adapters are active, refining the attention mechanisms to attend to glitch-specific features such as scattering arches or short-duration blips. We extract the final embeddings from the model's output layer, yielding a 768-dimensional vector for each glitch. To visualize the semantic structure of the glitch population, we apply a two-stage dimensionality reduction (Steps 3 \& 4). First, Principal Component Analysis (PCA) is applied to reduce the embedding to 128 dimensions, filtering out low-variance components that are often dominated by noise and redundancy, and making the subsequent manifold learning more stable and efficient. Finally, t-SNE is run on these PCA-compressed features to project them into a 3D latent space for visualization and clustering. \\

t-SNE is a powerful non-linear dimensionality reduction technique that is used to visualize the semantic structure of a high-dimensional feature space. Unlike linear methods, t-SNE excels at preserving the local structure of the data, meaning that glitches that are morphologically similar in the high-dimensional feature space are mapped to nearby points in the resulting three-dimensional latent space. This process makes it possible to visually inspect the data and identify distinct populations or clusters of glitch morphologies without requiring discrete, manual labels. \\


Compared to applying t-SNE directly to engineered time–frequency summaries, as in Ferreira and González (2025) \citep{Tabata_tSNE}, our framework first maps each glitch spectrogram into a compact, learned representation using the AST encoder. The large-scale pre-training of AST provides a strong inductive bias for spectro-temporal structure, which we then adapt to O3 glitches via LoRA, yielding embeddings that are more directly aligned with morphological similarity. This can improve separability and stability of glitch groupings relative to hand-engineered descriptors, especially when morphologies are subtle or heterogeneous. While both approaches inherit the known sensitivities of t-SNE (e.g., to hyperparameters and local density structure), the key distinction lies in the choice of input representation: Ferreira and González (2025) compress the data through engineered feature design (e.g., Q-transform/trigger summaries), whereas our method uses a learned embedding through a pre-trained transformer model. In practice, this shifts the main design burden from selecting handcrafted features to learning a representation that generalizes across glitch populations, while still requiring care in interpreting any t-SNE-based visualization. \\

\section{Results}

In this section, we describe tests performed using our method on O3 glitch data. The goal was to validate the effectiveness of our approach using quantitative clustering metrics and qualitative visualization. The primary experiment compared the clustering performance of two models: the vanilla AST encoder used off the shelf, without any exposure to LIGO glitch data, and our fine-tuned AST encoder, adapted to the O3 dataset using LoRA. In both cases, the extracted high-dimensional features were reduced using PCA and then projected with t-SNE for visualization. \\

A supplementary experiment assessed the model’s inherent ability to separate true GW signals from background noise without additional training (zero-shot anomaly detection). This was done by injecting two classes of BBH signals into O3 background noise: “vanilla” BBHs (total mass between 10 and 100 M$_{\odot}$) and short-duration intermediate-mass black hole (IMBH) binaries. For these tests, we validated our results against Gravity Spy classifications, which we treated as the “true” glitch labels. We also performed an agnostic clustering analysis using agglomerative clustering, similar to Ferreira and González (2025) \citep{Tabata_tSNE}, and cross-matched the resulting clusters with the Gravity Spy labels. We present the results of these tests below. \\

\subsection{Results with Gravity Spy as Reference Labels}

\begin{figure}
 \centering
        \includegraphics[width=0.7\textwidth]{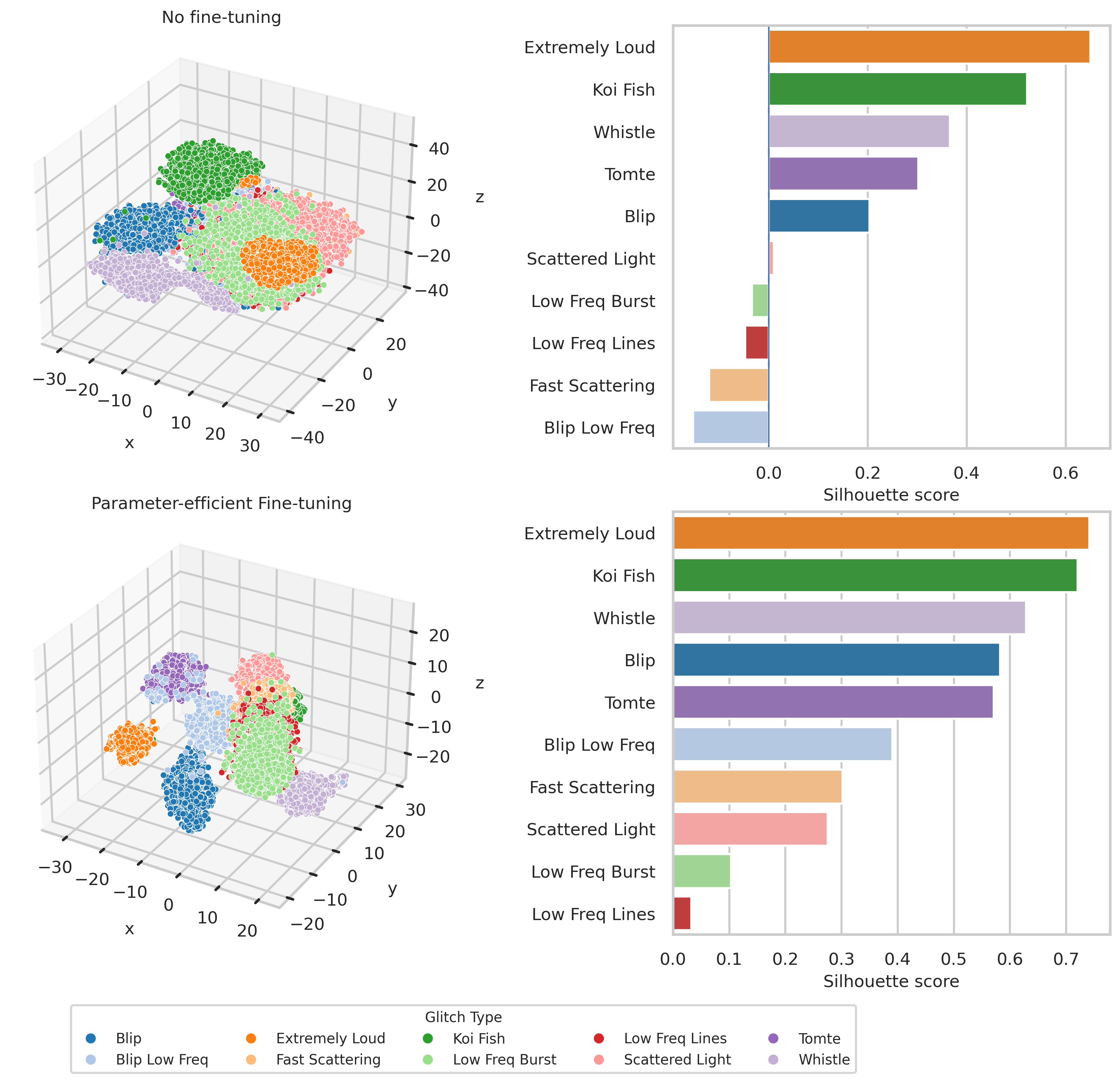}
 \caption{Top: t-SNE map (left) and Silhouette scores (right) of glitch embeddings obtained from off-the-shelf AST encoder. The embeddings were generated by an AST encoder with no exposure to glitch data during training. Botton: Same as the top figures, but with embeddings obtained from AST encoder fine-tuned using PEFT (LoRA) on O3 glitch data.}
\label{fig:fig3}
\end{figure}

Fig.~\ref{fig:fig3} shows the results of applying our method to 10 O3 glitch classes: Extremely Loud, Koi Fish, Whistle, Tomte, Blip, Scattered Light, Low-Frequency Burst, Low-Frequency Lines, Fast Scattering, and Blip Low Freq. We selected these 10 classes because they are common in O3, have high-confidence Gravity Spy labels, and span a wide range of time–frequency morphologies—including several low-frequency and Blip-like families that are known to be challenging and relevant for GW-search data quality—thereby providing a stringent and representative benchmark for evaluating embedding separability and clustering performance. \\

The left panels display 3D t-SNE projections of AST embeddings obtained using the off-the-shelf AST encoder (top; no fine-tuning on LIGO glitches) and after LoRA fine-tuning on O3 glitches (bottom). The right panels report the silhouette score for each glitch class in the two settings. The silhouette score measures how similar a data point is to its own cluster compared to other clusters. For any given data point $i$, the score $s(i)$ is calculated using two values: $a(i)$, the mean Euclidean distance between point $i$ and all other points in the same cluster, and $b(i)$, the mean Euclidean distance between point $i$ and the points in the nearest neighboring cluster. The expression is as follows:

\begin{equation}
    s(i)=\frac{b(i)-a(i)}{max(a(i),b(i))}
\end{equation}

A class-wise (or cluster-wise) Silhouette score is typically the average over all points in that class/cluster. Values closer to 1 indicate well-separated, compact clusters; values near 0 suggest overlapping clusters. Negative values indicate points that are, on average, closer to a different cluster than their assigned one. \\

The results show that fine-tuning substantially improves the organization of the AST feature space, but they also highlight that the off-the-shelf encoder already captures useful structure. Without any fine-tuning, the t-SNE projections reveal several partially coherent groupings, and the Silhouette scores indicate that a number of glitch classes (e.g., Extremely Loud, Koi Fish, Whistle, Tomte, and Blip) are already reasonably separable using generic acoustic representations. However, the off-the-shelf model struggles most for the low-frequency morphologies—particularly Low-frequency lines, Low-frequency burst, Fast scattering and Low-frequency blips -- which exhibit stronger overlap in the embedding space and correspondingly low (sometimes near-zero or negative) silhouette scores. This behavior is consistent with the expectation that domain mismatch is most severe for nonstationary, and low-frequency detector artifacts that don't have close analog in natural audio. \\

After LoRA fine-tuning on O3 glitches, the embedding space becomes markedly more class-discriminative: the t-SNE projections show tighter, more compact clusters with clearer boundaries, and the silhouette scores increase across essentially all classes, with the most pronounced gains occurring for the low-frequency glitch families that were previously poorly resolved. This improvement demonstrates that LoRA effectively adapts the AST encoder to LIGO-specific time–frequency morphology, refining its inductive bias so that subtle inter-class differences in detector noise are represented as separable directions in feature space. For fine-tuning AST, we used between 4,000 - 8,000 glitches per class drawn from Gravity Spy, restricting to events with classification probability $>$ 90\%, ensuring a high-purity training set. The per-class sample size is primarily constrained by the availability of high-confidence ($>$ 90\%) Gravity Spy labels and the desire to maintain a balanced, computationally tractable training set. We found that a few thousand examples per class were sufficient for LoRA to stabilize the embedding geometry, with additional samples providing diminishing returns while increasing training cost and class-imbalance risk. Finetuning the model took around 24 hrs on a single DGX A100 GPU. The evaluation time, including generation of the t-SNE plots took between 10-15 mins on the same resource. We performed filtering and whitening of the strain data before running it through AST, for which we used the relevant functions in the PyCBC package. Our code was written in PyTorch. \\

\begin{figure}
 \centering
        \includegraphics[width=0.7\textwidth]{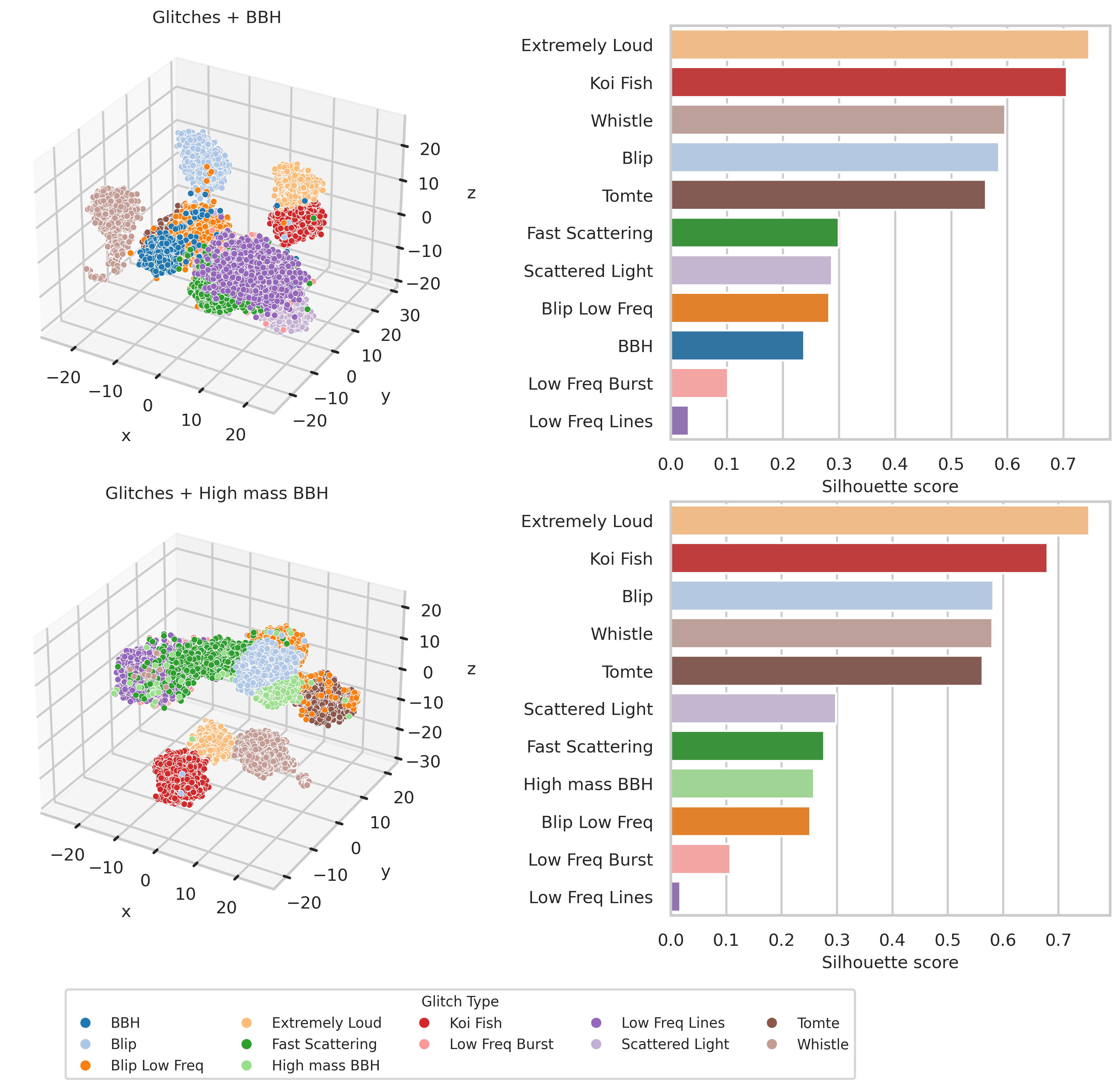}
 \caption{Top: t-SNE map (left) and Silhouette scores (right) of embeddings for standard BBH injections and 10 glitch classes. The embeddings were generated by an AST encoder fine-tuned exclusively on O3 glitches, with no exposure to GW data during training. Bottom: t-SNE map (left) and Silhouette scores (right) of high-mass BBH injections (total mass $\sim$ 100–1000 M$_{\odot}$) and glitch embeddings obtained using the same model.}
\label{fig:fig4}
\end{figure}

Figure~\ref{fig:fig4} illustrates results obtained from the glitch-tuned AST model when GW signals are introduced. The left panels show 3D t-SNE projections of the embedding space for the same O3 glitch classes as in Fig.~\ref{fig:fig3}, but now augmented with simulated BBH injections: the top row includes “vanilla” BBHs, while the bottom row includes short-duration, high-mass BBHs in the IMBH regime. The corresponding right panels report class-wise Silhouette scores for the joint datasets in each case. In both experiments, the injected GW population forms a relatively compact cluster that is generally displaced from the main regions occupied by the instrument glitch classes, though some overlap and boundary mixing remain. This separation is particularly notable because the model was not fine-tuned on GW signals at all -- the encoder was adapted only using glitch data -- yet the learned representation still places astrophysical waveforms in regions of feature space that are distinct from known noise morphologies. The silhouette scores are consistent with this qualitative picture: most glitch classes remain well-resolved while the GW class achieves a positive Silhouette value, indicating that it is closer (on average) to its own cluster than to any glitch cluster. Importantly, the same behavior holds for the more challenging IMBH-like injections: despite their short duration and potential morphological similarity to some transient glitches, the high-mass BBHs remain mostly separated from the noise clusters, with some scatter around Fast scattering and Low-frequency line clusters. This capability is useful in practice as an `anomaly detector' because it enables signal-agnostic candidate isolation. During live observing runs, the model can flag events that lie outside the manifold of known glitch morphologies as anomalous without requiring supervised training. \\

\subsection{Agglomerative Clustering on AST embeddings}

\begin{figure}
 \centering
        \includegraphics[width=1.0\textwidth]{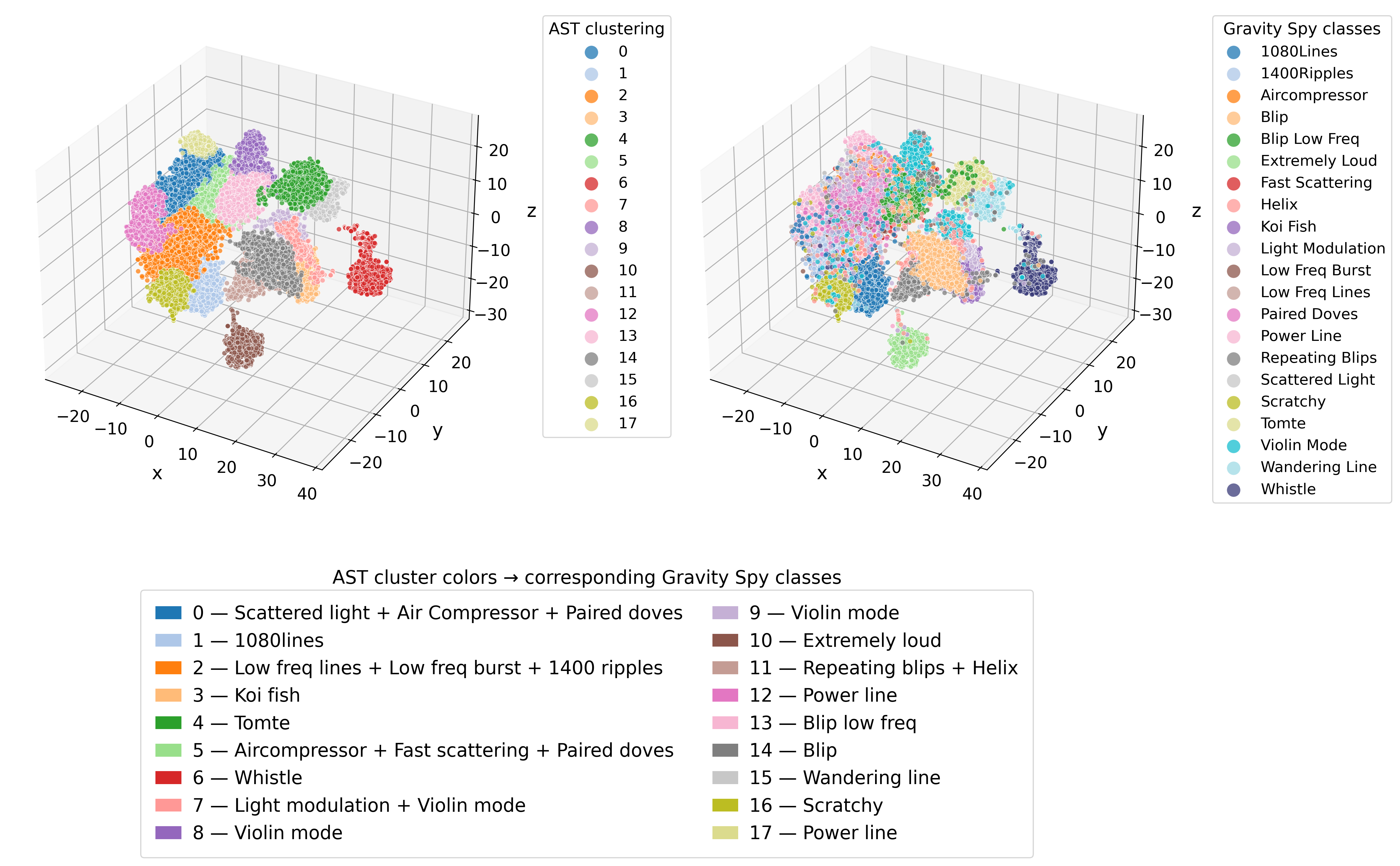}
 \caption{\textit{Top left:} 3D t-SNE visualization of AST embeddings for 21 glitch types across all three observing runs, colored by unsupervised agglomerative (hierarchical) clustering applied in the 2D/3D t-SNE space following the procedure of Ferreira \& González (2025) \citep{Tabata_tSNE}. \textit{Top right:} The same embeddings colored by Gravity Spy classifications. \textit{Bottom:} Cross-match between agglomerative cluster labels and the dominant Gravity Spy class(es) present in each cluster. Our model identifies 18 clusters (labels 0–17), reflecting mergers of morphologically similar or overlapping Gravity Spy categories (e.g., related low-frequency families), and splits of some labeled classes into sub-populations when the embedding space resolves distinct modes.}
\label{fig:fig5}
\end{figure}

Fig.~\ref{fig:fig5} shows a 3D t-SNE visualization of AST embedding vectors computed for 21 glitch types across all three observing runs. We expanded from 10 to 21 classes for this test to probe robustness under a more realistic and heterogeneous glitch population and to assess stability and substructure across observing runs. Following Ferreira \& González (2025) \citep{Tabata_tSNE}, we apply agglomerative (hierarchical) clustering using Ward's method to assign each point to a cluster. In this method, clusters are built by first treating each data point as its own cluster and then iteratively merging pairs to minimize the total in-cluster variance (sum of squared Euclidean distances to cluster centroids). The optimal number of clusters is determined automatically by maximizing the mean Silhouette scores across all the clusters. \\

In the left panel, the t-SNE points are colored by the resulting Ward agglomerative labels. The method recovers 18 clusters (labels 0–17) rather than 21, indicating that in this AST feature space, some Gravity Spy glitch morphologies are not fully separable, while others contain internal substructure that the clustering algorithm treats as distinct groups. In the right panel, the same t-SNE embedding is colored by the Gravity Spy classifications, and the mapping table below cross-matches each agglomerative cluster to the dominant Gravity Spy label(s). \\

Overall, the agreement is strong, as many clusters correspond approximately one-to-one with a single Gravity Spy class (e.g., clear, compact regions associated with classes such as Koi Fish, Tomte, Whistle, Extremely Loud, Blip, and Wandering Line). At the same time, several clusters naturally merge multiple Gravity Spy labels (e.g., low-frequency families being grouped together, or morphologically related classes such as Repeating Blips and Helix appearing in the same region), consistent with the fact that these glitch categories can share overlapping time–frequency morphologies. Conversely, some Gravity Spy classes appear split across multiple agglomerative clusters (e.g., Violin Mode and Power Line), suggesting the presence of distinct sub-populations within the same label in the learned representation. This likely reflects evolution in characteristics of these glitch morphologies across observing runs, such that the clustering algorithm associates subtle but systematic feature differences with distinct groups in the embedding space. \\

\subsection{Generalization test using Omicron triggers from O4}

\begin{figure}
 \centering
        \includegraphics[width=0.8\textwidth]{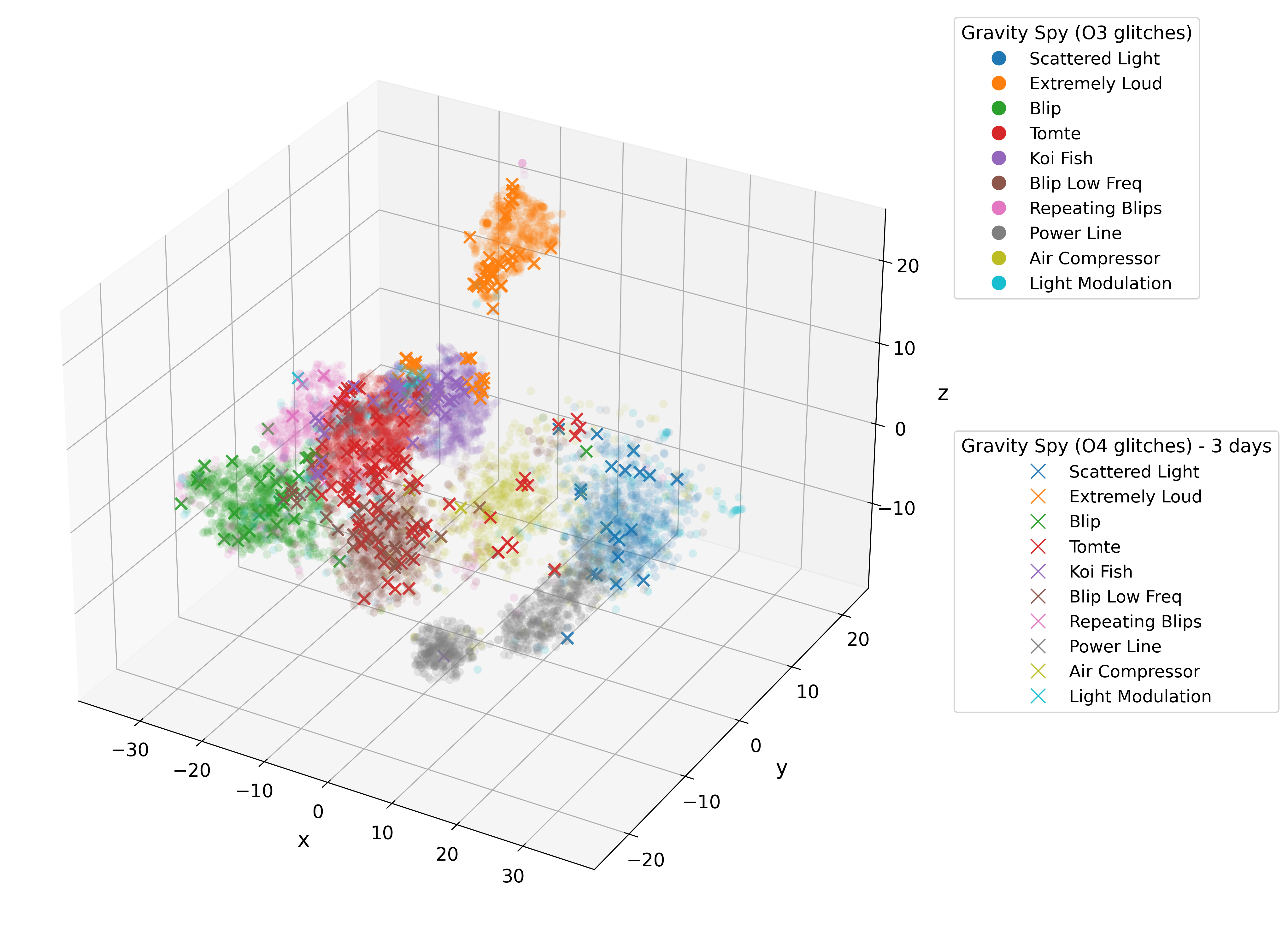}
 \caption{Projection of O4 Omicron triggers into the O3-trained AST embedding space. Semi-transparent point clouds show 3D t-SNE projections of AST embeddings for selected O3 glitch classes (colored by their Gravity Spy labels), forming the reference clusters learned from O3 data. Cross markers show Omicron triggers from a 4-day segment of O4 data (GPS 1372723218–1373068817), colored by their Gravity Spy classifications in O4. The extent to which O4 crosses overlap the corresponding O3 clusters provides a cross-run generalization test of the learned morphology representation.}
\label{fig:fig6}
\end{figure}

Fig.~\ref{fig:fig6} presents a cross-run generalization test in which we project short segments of O4 Omicron triggers into the O3-trained AST embedding space and examine whether they fall into the same regions occupied by the corresponding O3 glitch populations. Omicron is an online, low-latency transient-finding algorithm used in LVK detector characterization that identifies short-duration excess-power events in time–frequency space and records their properties (e.g., central time, frequency, quality factor, and SNR) \citep{Omicron}. Here, we select Omicron triggers from a 4-day span of O4 data (GPS times between 1372723218 and 1373068817) and assign each trigger a glitch label using the Gravity Spy classifier. Over the 4-day segment, 587 Omicron triggers and 578 Gravity Spy events were found in L1. By cross-matching the trigger GPS times, 477 events were identified to be common to both pipelines and used in our analysis. \\

In the plot, the semi-transparent point clouds show the AST embedding clusters derived previously from O3 glitch examples (colored by their O3 Gravity Spy class). The cross markers show the O4 Omicron triggers, colored by their O4 Gravity Spy classifications. Because the AST encoder was tuned using O3 glitches, this provides a direct test of whether O4 triggers—detected independently by Omicron and labeled by Gravity Spy—map onto the same learned morphology manifold. \\

Overall, for most classes, the O4 Omicron triggers lie predominantly within (or very near) the corresponding O3 clusters. This indicates that the representation learned from O3 captures stable, reusable morphological features that persist into O4. At the same time, some glitches like Tomte show a noticeable spread of O4 triggers around or partially outside the O3 regions. This dispersion can signify few non-exclusive effects like morphological evolution of a glitch family between O3 and O4 due to changes in detector configuration or increased intra-class diversity in O4. It could also indicate label ambiguity in O4 classifications, particularly for morphologically similar transient families, or the fact that Omicron triggers include a wide range of trigger strengths and may capture borderline events that sit between canonical classes. We plan to explore this further over longer random stretches of O4 data in our future work. \\

This test demonstrates the utility of the learned embedding for continuous detector-characterization monitoring since triggers that remain inside established clusters can be automatically associated with known glitch families, while triggers that systematically drift or populate new regions can be flagged as potential distribution shift or emergent glitch morphologies. This is especially valuable in low latency, where rapidly identifying “known” versus “novel” instrumental behavior can help prioritize follow-up and reduce false alarms during live observing. \\

\section{Discussion}

We presented a cross-domain framework for GW detector characterization that leverages the inductive bias of large, pre-trained audio transformers. Our method treats time-frequency representations of LIGO strain as an audio understanding problem: we convert short strain segments into log-mel spectrograms, embed them with the Audio Spectrogram Transformer (AST), and analyze the resulting representation space using dimensionality reduction and clustering. To adapt AST to interferometer noise morphology without expensive full fine-tuning, we employ LoRA, enabling parameter-efficient specialization to O3 glitch data while preserving the general audio-based priors learned from large-scale pre-training. \\

Across multiple complementary tests, we find that AST embeddings provide a semantically meaningful representation of glitch morphology, and that LoRA fine-tuning substantially sharpens this representation. Using Gravity Spy labels as reference classes, the off-the-shelf AST encoder already separates several high-SNR, visually distinct glitch families reasonably well, indicating that generic audio pre-training transfers non-trivially to GW detector artifacts. After LoRA adaptation, the embedding space becomes more class-discriminative, with tighter and more separated groupings and improved class-wise clustering metrics, particularly for low-frequency and morphologically heterogeneous glitch families. Ultimately, this framework offers a robust, data-efficient pathway for discovering new, anomalous transients, and classifying complex noise artifacts in the era of next-generation detectors. \\

A critical finding is the framework's potential for signal-agnostic anomaly detection. When simulated GW signals are embedded using a model fine-tuned exclusively on glitches (with no exposure to GW waveforms), the astrophysical populations occupy a compact region distinct from the dominant glitch clusters. Crucially, this separation persists even for short-duration, high-mass (IMBH-regime) injections that morphologically resemble transient glitches. This suggests a practical route to low-latency implementation that does not require supervised training on an exhaustive set of waveform families. Furthermore, our unsupervised population study (via agglomerative clustering) revealed both mergers of morphologically similar Gravity Spy classes and splits consistent with intra-class diversity. Finally, a cross-run generalization test projecting O4 Omicron triggers into the O3-trained embedding space indicates a potential distribution shift for Tomte-like triggers, plausibly driven by changes in detector configuration or labeling ambiguity in low-latency classifications. \\

These results support a workflow where known glitch families are monitored by tracking trigger density within established embedding regions, while events populating new regions or exhibiting systematic drift are prioritized for rapid follow-up. Because the model is adapted via LoRA using modest, high-purity labeled sets, it allows for rapid re-tuning as detector conditions evolve, effectively mitigating the "label bottleneck" inherent to supervised-from-scratch CNNs. Looking ahead, we plan to expand injection studies to broader waveform families (including precession, eccentricity, and higher modes) and to glitch-signal overlap scenarios. Finally, we will investigate continual-learning strategies, such as incremental LoRA adapters, to update the representation as new glitch morphologies emerge while preserving performance on previously learned populations. \\

\begin{acknowledgments}
The authors would like to thank Derek Davis and Debasmita Nandi for helpful comments and suggestions. This research was undertaken with the support of compute grant and resources, particularly the DGX A100 AI Computing Server, offered by the Vanderbilt Data Science Institute (DSI) located at Vanderbilt University, USA. This research used data obtained from the Gravitational Wave Open Science Center (https://www.gw-openscience.org), a service of LIGO Laboratory, the LIGO Scientific Collaboration and the Virgo Collaboration. LIGO is funded by the U.S. National Science Foundation. Virgo is funded by the French Centre National de Recherche Scientifique (CNRS), the Italian Istituto Nazionale della Fisica Nucleare (INFN) and the Dutch Nikhef, with contributions by Polish and Hungarian institutes. This material is based upon work supported by NSF's LIGO Laboratory which is a major facility fully funded by the National Science Foundation.
\end{acknowledgments}

\bibliography{sample701}{}
\bibliographystyle{aasjournalv7}



\end{document}